\documentclass[aps,nofootinbib,twocolumn,showpacs,pra,10pt,floatfix]{revtex4-1}
\usepackage{amsfonts}
\usepackage{amsmath}
\usepackage{amssymb}
\usepackage{graphicx}

\DeclareMathOperator{\Trace}{Tr}

\usepackage{color}

\begin{document}

\title{Four-boson bound states from a functional 
renormalisation group}
\author{Benjam\'in Jaramillo \'Avila and Michael C. Birse}
\affiliation{Theoretical Physics Division, School of Physics and Astronomy,\\
The University of Manchester, Manchester, M13 9PL, UK\\}

\begin{abstract}
We use the functional renormalisation group to study the spectrum of 
three- and four-body states in bosonic systems around the unitary limit.
Our effective action includes all energy-independent contact interactions in 
the four-atom sector and we introduce a running trimer field to eliminate 
couplings that involve the atom-atom-dimer channel. The results show
qualitatively similar behaviour to those from exact approaches. The truncated 
action we use leads to overbinding of the two four-body states seen in those treatments. It also generates a third state, although only for a very narrow 
range of two-body scattering lengths.
\end{abstract}

\maketitle

\vskip 10pt

Systems where the two-body scattering is close to the unitary limit display
universal features that are independent of the underlying dynamics. This physics
has been realised by ultra-cold atoms in traps, where the scattering length can 
tuned using Feshbach resonances \cite{cgjt10,fzbhng11,hp11}. It may also be 
relevant to low-energy aspects of few-nucleon systems \cite{bvk02,hp10}.

A key feature of this limit for bosonic systems (and also for fermionic ones with
at least three species of particle) is the Efimov effect \cite{efim70,efim71}.
This is the appearance of an infinite tower of three-body bound states, with energies in a constant ratio. As a result the scale invariance of the unitary limit is
broken leaving only a discrete symmetry where momenta are scaled by a factor of 22.7.
This physics feeds through to systems with more particles. Two four-body bound 
states (or, strictly speaking, narrow resonances) 
have been found just below each three-body state \cite{hp07,vsdg09,delt10,delt11a}. 
More recently these analyses have been extended to states of up 16 
bosons \cite{gkv12,ktg14}.

The scaling behaviour of the three-boson system has been analysed using the
renormalisation group (RG) and the Efimov effect has been shown to result from
a limit cycle in the flow of the leading three-body contact interaction
\cite{bhvk99,bhvk00,bb05,hp11}. However it has proved impossible so far to extend 
these rigorous RG approaches to systems of four or more particles. A promising
alternative tool for exploring scaling in these systems is the functional renormalisation group (FRG), particularly the version based on the
Legendre-transformed effective action developed by Wetterich and co-workers
\cite{wett93,btw02,paw07,dfgpw10}. Although in principle the method is exact, 
practical applications rely on truncations of the action to a finite number of 
terms. Nonetheless, even with fairly simple truncations, it has yielded useful 
results across a wide range of areas \cite{dfgpw10,delam12,gies12}.

In previous work \cite{jb13}, we applied the FRG to the four-boson problem,
using it to determine the values of the two-body scattering length for which
the four-body bound states appear at zero energy. These correspond to 
points at which resonant recombination occurs in trapped cold-atom systems.
Our results for these scattering lengths are in good agreement with those 
from exact few-body calculations \cite{vsdg09,delt10,delt12}
and experiment \cite{f09}. 
That work showed that it was important to keep all zero-derivative contact 
terms, including ones where dimers break up into pairs of atoms 
\cite{sm10,bkw11}. It also demonstrated that energy dependence in the trimer
channel was needed to generate four-body bound states. That energy dependence
was described through the introduction of a trimer field \cite{fmsw09,sm10}.

The results of Ref.~\cite{jb13} were restricted to zero-energy physics;
here we extend that analysis to nonzero energy in order to access 
the spectrum of four-body states. Our starting point is a minor modification 
of the effective action used in that work:
\begin{widetext}
\begin{eqnarray}\label{eq:running:action}
\Gamma_k[\psi,\psi^*,\phi,\phi^*,\chi,\chi^*]
&=&\int{\rm d}^4x\,\Biggl[
\psi^*\left({\rm i}\,\partial_0+E_a+\frac{\nabla^2}{2m}\right)\psi
+Z_d\,\phi^*\left({\rm i}\,\partial_0+\frac{\nabla^2}{4m}\right)\phi
+Z_t\,\chi^*\left({\rm i}\,\partial_0+\frac{\nabla^2}{6m}\right)\chi\cr
\noalign{\vspace{5pt}}
&&\qquad\qquad-u_d\phi^*\phi-u_t\chi^*\chi
-\frac{g}{2}\bigl(\phi^*\psi\psi+\psi^*\psi^*\phi\bigr)
-h\bigl(\chi^*\phi\psi+\phi^*\psi^*\chi\bigr)-\lambda\,\phi^*\psi^*\phi\psi\cr
\noalign{\vspace{5pt}}
&&\qquad\qquad-\frac{u_{dd}}{2}\bigl(\phi^*\phi\bigr)^2
-\frac{v_d}{4}\bigl(\phi^*\phi^*\phi\psi\psi+\phi^*\psi^*\psi^*\phi\phi\bigr)
-\frac{w}{4}\phi^*\psi^*\psi^*\phi\psi\psi\cr
&&\qquad\qquad\null -u_{tt}\, \chi^*\psi^*\chi\psi
-\frac{u_{dt}}{2}\bigl(\phi^*\phi^*\chi\psi+\chi^*\psi^*\phi\phi\bigr)
-\frac{v_t}{2}\bigl(\phi^*\psi^*\psi^*\chi\psi+\chi^*\psi^*\phi\psi\psi\bigr)
\Biggr].
\end{eqnarray}
\end{widetext}
Here $\psi$, $\phi$ and $\chi$ are the fields corresponding to atoms (A), dimers
(D) and trimers (T) respectively.
This action is supplemented by regulators for each field that suppress the 
contributions of fluctuations with momenta below some scale $k$. As in our 
previous work, we take the form suggested by Litim \cite{lit01} for the 
regulators $R_{a,d,t}(q,k)$. The renormalisation group (RG) equation describes 
the evolution of the effective action as $k$ is lowered and more and more modes 
are integrated out until, in the limit $k\rightarrow 0$, the physical effective 
action is reached.

We work with a trimer field that runs with renormalisation scale
\cite{fmsw09,sm10}. This allows us to absorb the interaction terms that contain 
the AAD ``breakup" channel, leaving only the DD and AT channels that are present 
in the Faddeev-Yakubovsky equations \cite{yak67}. 
The resulting RG equation for the effective action has the form
\begin{eqnarray}\label{eq:FRG}
\partial_k\Gamma&=&-\frac{\mbox{i}}{2}\,\Trace \left[(\partial_k{\mathbf R})\,
\left(({\boldsymbol \Gamma}^{(2)}-{\mathbf R})^{-1}\right)\right]\cr
\noalign{\vspace{5pt}}
&&+\frac{\delta\Gamma}{\delta\chi}\cdot\partial_k\chi
+\frac{\delta\Gamma}{\delta\chi^*}\cdot\partial_k\chi^*\,,
\end{eqnarray}
where the final terms are generated by the running of the trimer field 
\cite{gw02,paw07}. Floerchinger and Wetterich \cite{fw09} have proposed a more
complicated version of the flow equation for a running field, which they derive 
from a change of field variables in the functional integral. The version used
here can be thought of as arising instead from a shift of the expansion point
of the field, in order to cancel certain terms in the effective action. This 
is analogous to the shift in the expansion point often used to treat systems 
where a symmetry is spontaneously broken (see, for example, 
Refs.~\cite{btw02,bkmw05,dgpw07a}).

Ordinary differential equations describing the running of each of the couplings
are obtained by expanding both sides of Eq.~(\ref{eq:FRG}) in powers of the 
fields. More details of these equations can be found in Ref.~\cite{jb13} 
(see also Refs.~\cite{sm10,bkw11}).

The one difference between the action (\ref{eq:running:action}) and the one in 
Ref.~\cite{jb13} is that here we add an external 
energy $E_a=-\gamma^2/2m$ flowing along each atom line. In effect this shifts 
the expansion point for the part of action describing a system of $n$ atoms from 
zero energy to $nE_A$. The results would be independent of this expansion point if
we worked with the full, untruncated effective action. In practice, 
some artefacts of the truncation to a set of local terms are visible in our 
results and provide some indication of the errors introduced by it.

Note that all the parameters are implicitly functions of the external
energy $E_a$  and that we have chosen not to subtract a term $2Z_d E_a$ from
the dimer self-energy $u_d$. The condition for a dimer bound state is 
therefore $u_d=0$ (rather than $2E_a=u_d/Z_d$). Similarly, the condition
for a trimer bound state is $u_t=0$.

We start the evolution at some large scale $K$ where only the 
atom fields are dynamical ($Z_\phi=Z_\chi=0$) and $u_d(K)$ is fixed so 
that $g^2/u_d(k=0)$ gives the physical atom-atom scattering length 
$a$ \cite{bkmw05,dgpw07a}. For a fixed value of the energy $E_a$, we scan through
values of $a$ to identify the ones at which the three-body energy, $u_t$, vanishes,
or the four-body couplings, $u_{dd}$ etc., diverge in the physical limit. These
correspond to points at which an $n$-body bound state has energy $nE_a$. 

The evolution in the three-body sector shows the periodic behaviour expected 
as result of the Efimov effect \cite{efim70,efim71}. However, the truncation of 
the action means that the momentum scaling factor is somewhat larger than 
the true value, $29.8$ rather than $22.7$ \cite{fmsw09,sm10}.

Another consequence of truncating the energy dependence in the action is that 
some thresholds do not not appear in the correct places. For example, the
atom-dimer threshold in the trimer propagator should be where the total energy 
in the three-body system is equal to the dimer binding energy, $3E_a=-1/ma^2$.
Within our truncation, it appears where the atom-dimer energy
$E_a+u_d/Z_d$ vanishes. Using the expressions for the full two-body self 
energy at $k=0$ \cite{b08},
\begin{eqnarray}
u_d(E_a,0) &=& \frac{g^2m}{4\pi^2}\left({\rm i}\sqrt{2mE_a}
+\frac{1}{a}\right),\\
Z_d(E_a,0) &=& \frac{1}{2}\,\frac{\partial u_d}{\partial E_a} 
=\frac{g^2m^2}{8\pi^2}\,\frac{{\rm i}}{\sqrt{2mE_a}}\,,
\end{eqnarray}
we find that this condition is satisfied for 
\begin{equation}
E_a=-\,\frac{1}{2m}\left(\frac{4}{5a}\right)^2.
\end{equation}
This effective threshold corresponds to a total three-body energy that differs by 
a factor of $24/25$ from the correct one. A similar issue arises for the atom-trimer
threshold in the four-body sector.

\begin{figure} [h]
\includegraphics[width=\columnwidth]{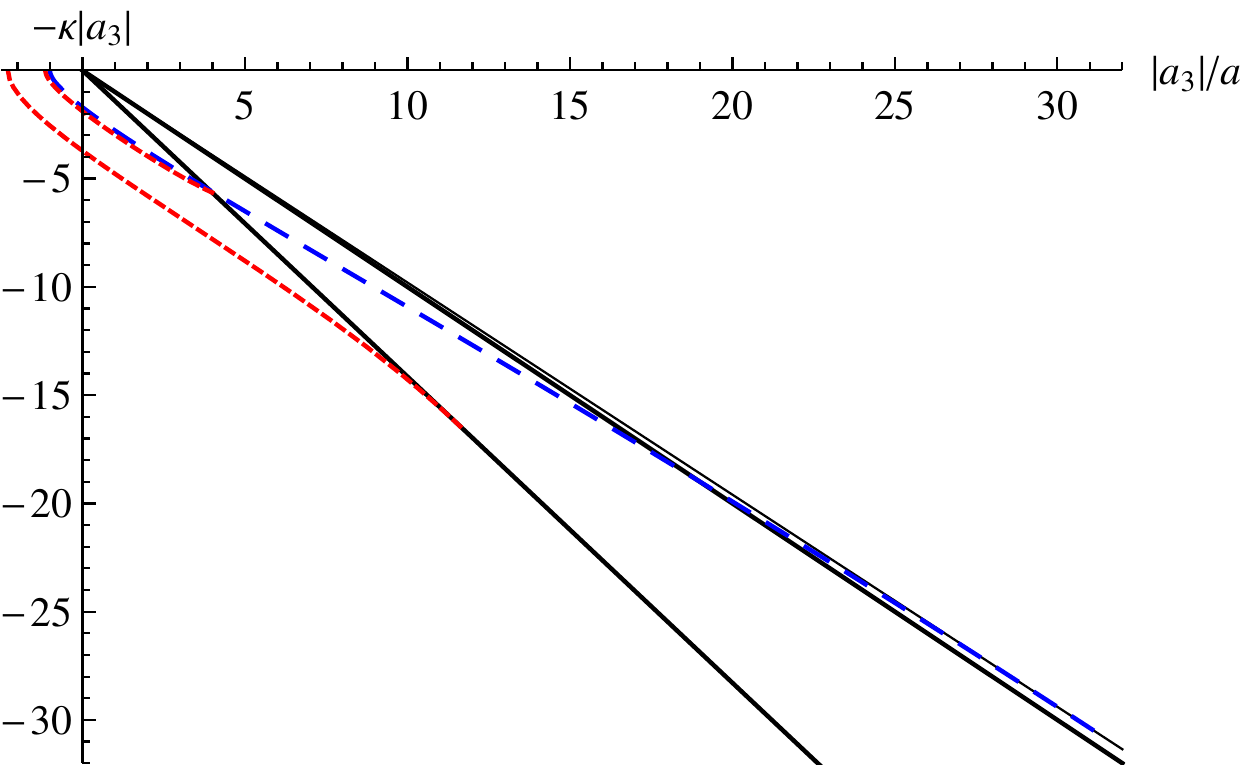}
\caption{
\label{fig:ef_full}
Three- and four-body bound states for one Efimov cycle. The binding momenta 
$-\kappa|a_3|$ are plotted against the inverse atom-atom scattering length, 
$|a_3|/a$. The thick solid lines show the dimer bound state and the DD
threshold. The long-dashed blue line shows the trimer bound state, and 
the short-dashed red lines the tetramer bound states. The thin solid line shows 
the effective AD threshold for our truncation.}
\end{figure}

As already found in our study of this system at zero energy, our approach 
leads to three four-body bound states in each Efimov cycle \cite{jb13}. 
This is one more than are seen in exact solutions of the four-body problem
\cite{hp07,vsdg09,delt10,delt11b,delt12} and is likely to reflect the fact
that our Efimov cycles are too long. Our results for the binding energies of 
the three- and four-body states in a single cycle are presented in Figs.~1 and 2. 
The plots show the energies, expressed in the form of binding momenta,
$\kappa=\sqrt{-n\,mE_a}$, as functions of the inverse scattering length $1/a$. 
All quantities have been expressed in units of $1/|a_3|$, where $a_3$ is the 
atom-atom scattering length at which the three-body state becomes unbound unto 
three atoms. 

\begin{figure} [ht]
\includegraphics[width=\columnwidth]{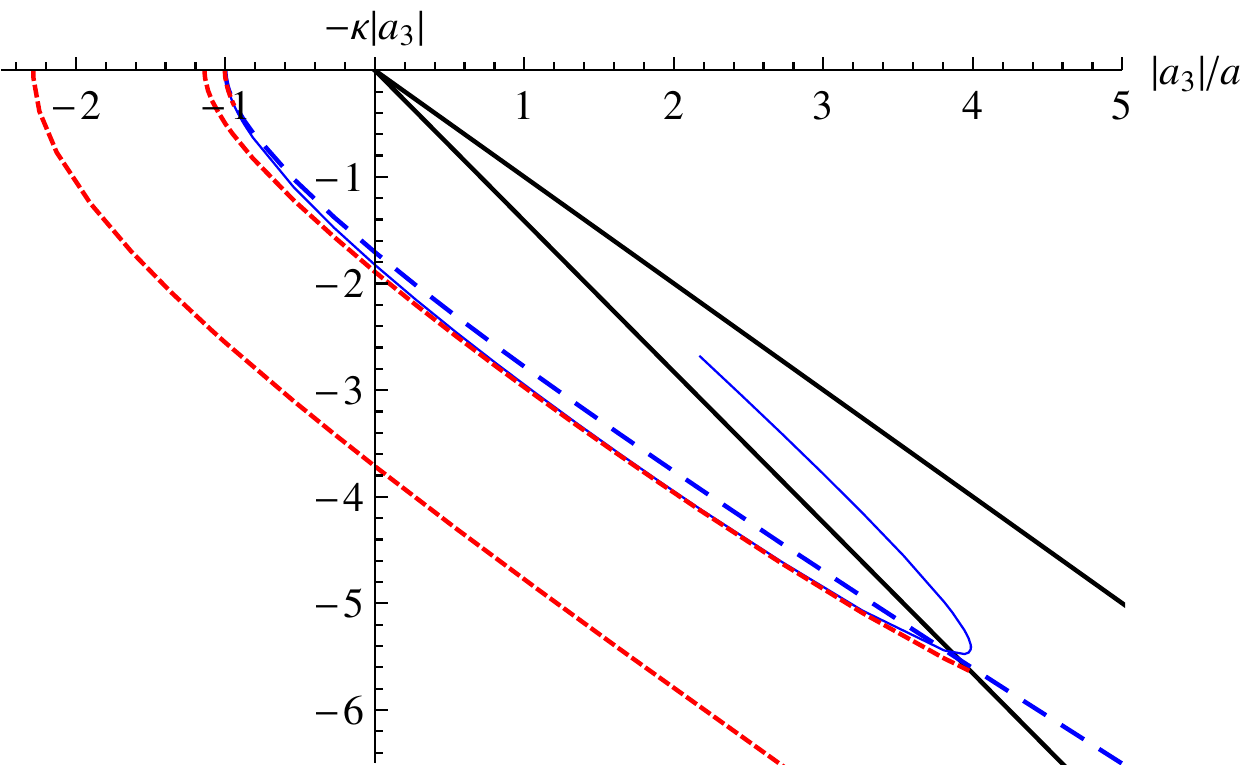}
\caption{
\label{fig:ef_zoom_1}
Expanded view of the three- and four-body bound states with smaller energies.
The notation is as in Fig.~1 except for the thin blue line which here shows
the effective AT threshold.}
\end{figure}

The three-body state is bound for atom-atom scattering lengths 
$|a_3|/a>-1$. Its binding momentum in the unitary limit is $\kappa_3|a_3|=1.71$,
which should be compared with the exact result $\kappa_3|a_3|=1.5077$
from Deltuva's solutions of the Faddeev equation \cite{delt12}. Eventually the 
trimer becomes unbound against decay into the AD channel. It is difficult
to determine precisely the point at which this happens in our approach because,
as discussed above, the threshold is slightly displaced in our truncation.
Since the curves for the dimer and trimer binding energies cross at a very 
shallow angle, the small error in the threshold position has a 
disproportionately large effect on the point at which the trimer becomes 
unbound. Its energy passes through the dimer binding energy at 
$|a_3|/a\simeq +19$, but does not reach the effective threshold until 
$|a_3|/a=+31.1$. For comparison, Deltuva's exact solutions give 
$|a_3|/a=+21.30$ \cite{delt11b,delt12}.

The tetramer states become bound with respect to the four-atom channel at
$|a_3|/a=-2.29$, $-1.14$ and $-1.003$ \cite{jb13}. For comparison, 
the two states seen in exact treatments appear at $|a_3|/a=-2.351$ 
and $-1.096$ \cite{delt12}. Only two of our states persist to the unitary 
limit, where their binding momenta are $\kappa|a_3|=3.71$ and 1.89. 
Relative to the three-body state, these are $\kappa/\kappa_3=2.18$ and 
1.11, and so both are
overbound compared to the exact results, 2.147 and 1.0011 \cite{delt10}. 
The third state is very weakly bound and so is almost invisible in the plots. 
It is bound only within a very narrow range of atom-atom scattering lengths, 
hitting the effective AT threshold (which lies slightly below 
the actual trimer binding energy) at $|a_3|/a=-0.947$.

The deepest tetramer becomes unbound when it reaches the DD threshold
at $|a_3|/a=+11.75$, which should be compared with $+9.700$ from 
the numbers in Refs.~\cite{delt11b,delt12}. The second tetramer state 
remains bound with respect to the trimer until it hits that threshold at 
$|a_3|/a=+3.99$, to be compared with $+3.140$. This behaviour contrasts with
what is seen in exact treatments \cite{delt11a,delt13}, where the state is 
a virtual one over much of the region between $1/a=0$ and the point where the
trimer crosses the dimer-dimer threshold. However it is perhaps worth pointing 
out that the state lies very close to the effective AT threshold over 
much of this range.

The energy spectrum shows no sign of the super-Efimov behaviour -- an infinite 
tower of states in a double exponential pattern \cite{mns13} -- seen during the
evolution in the regime where the cutoff scale much larger than $1/a$ and $\gamma$.
This is consistent with the theorem of Amado and Greenwood, which states that there cannot be an infinite number of tetramers with an accumulation point at zero 
energy \cite{ag73}. We also see no sign of the AAD Efimov effect predicted by 
Braaten and Hammer \cite{bh06} in region where the trimer crosses the DD 
threshold. This may be another artefact of our truncation since just beyond 
this point our effective atom-trimer effective AT threshold starts to show 
unphysical behaviour, bending back on itself instead of following the
trimer binding energy (see Fig.~2).

In this work we have extended our previous FRG approach \cite{jb13} to study
the spectrum of three- and four-body states in systems close to the unitary limit.
In order to include energy dependence in the three-body sector we introduce
a trimer field. We work with a truncated action that retains all 
energy-independent contact interactions in the four-body sector. We allow 
the trimer field to run with our cutoff in order to eliminate couplings 
that involve the AAD "breakup" channel and hence to generate equations with 
an analogous structure to the Faddeev-Yakubovsky equations.

Our results show similar qualitative behaviour to those obtained from exact 
treatments of the few-body systems. In many cases we even get quantitative 
agreement at the 20\% level or better. The main exceptions are the length of the 
Efimov cycle, which is about 30\% too large, and the overbinding of the four-body 
states which leads to the appearance of a third tetramer, if only in a very narrow
window. This reinforces the importance of keeping all four-body contact 
interactions and of including energy dependence in the two- and three-body 
subsystems. 

Improving on these results will require the use of truncations
that include more of this energy dependence, though higher-order terms in the
propagators and energy- or momentum-dependent couplings. With such 
improvements, it will be interesting to explore whether the FRG can also
describe systems with more particles and strongly interacting bosonic matter.

\begin{acknowledgments}
BJA acknowledges support from CONACyT. This work was also supported in part by
the UK STFC under grant ST/J000159/1.
MCB is grateful to the Institute for Nuclear Theory, Seattle for its hospitality
and the US Department of Energy for partial support. We would also like to thank 
the participants at the 2014 workshop on Few-Body Universality in Atomic and 
Nuclear Physics, especially D. Blume, E. Braaten, M. Gattobigio, S. Moroz and
D. Phillips, for discussions that prompted this work.

\end{acknowledgments}

\end{document}